\documentclass[prd,twocolumn,superscriptaddress,altaffilletter,amssymb,showpacs,nofootinbib]{revtex4}
\usepackage[dvips]{graphicx}
\usepackage{amsmath}
\newcommand{\be}{\begin{equation}}
\newcommand{\ee}{\end{equation}}
\newcommand{\ben}{\begin{eqnarray}}
\newcommand{\een}{\end{eqnarray}}
\newcommand{\vphi}{\varphi}
\newcommand{\ds}{\displaystyle}

\begin{document}

\title{Quintom cosmologies with arbitrary potentials}
\author{Ruth Lazkoz}
\email{ruth.lazkoz@ehu.es}
\affiliation{Fisika Teorikoa, Zientzia eta Teknologia Fakultatea, Euskal Herriko Unibertsitatea, 644 Posta Kutxatila, 48080 Bilbao, Spain}
\author{Genly Le\'on}\email{genly@uclv.edu.cu}\affiliation{Universidad Central de Las Villas, Santa Clara CP 54830, Cuba}
\author{Israel Quiros}\email{israel@uclv.edu.cu}\affiliation{Universidad Central de Las Villas, Santa Clara CP 54830, Cuba}
\date{\today}
\begin{abstract}
In this paper we investigate quintom cosmologies with arbitrary potentials from the dynamical systems perspective. The dynamical systems analysis is complete in the sense that it includes the asymptotic regime where both scalar fields diverge, which proves to be particularly relevant in connection with the existence of tracking phases. The results of the present study indicate that the existence of phantom attractors is not generic: for quintom models there may exist either de Sitter attractors associated with the saddle points of the potential, or tracking attractors in the asymptotic regime where the scalar fields diverge.
\end{abstract}
\maketitle

\section{Introduction}

Observational cosmology supports the claim that our universe is currently undergoing accelerated expansion \cite{sn,cmb,sdss}. Within the framework of Einstein's theory of general relativity, the arguably only possible explanation to this phenomenon relies on the existence of an enigmatic component of the universe acting as a repulsive force, the so called dark energy (DE).

Dark energy is one of the most debated topics in the era of precision cosmology (see \cite{Sahni,Copeland:2006wr} for recent reviews). A cosmological constant (CC) is the simplest candidate to account for the DE, but this proposal suffers from the well-known cosmological constant problem according to which there exists a (yet) unexplained extraordinary discrepancy (of about 120 orders of magnitude) between the (cosmologically) observed value of the CC and the value predicted by quantum field theory. For this reason, alternative routes has been proposed support
(see again \cite{Copeland:2006wr}): quintessence, phantom dark energy, tachyon fields, Chaplygin gas, and recently, quintom dark energy. The latter alternative has aroused interest due to a debate about the possibility that the DE equation of state (EOS) parameter $w(z)$ crosses the phantom divide line $w=-1$ at recent redshifts.\footnote{The quantum stability of a $w< -1$ phase of cosmic acceleration was analyzed in \cite{Kahya}.} If this possibility were eventually corroborated, then one could think of two possible explanations/implications: either the DE consists of multiple components with (at least) a non-canonical phantom component, or general relativity has to be modified on cosmological scales \cite{nesseris}. The possibility to deal with several scalar fields is supported, besides, by the effective (low-energy) theory of the bosonic string where several scalar fields like the dilaton, the axion and other moduli fields can be found \cite{dwands}.

If dark energy becomes phantom-like (phantom dark energy has an equation of state (EOS) parameter $w$ smaller that $-1$) at recent times then, typically, our universe (if ever expanding) can evolve to a catastrophic singularity in the near future, characterized by divergences of the scale factor, the Hubble expansion scalar and its time derivative. It is the case if the dominant fractional energy density is phantom like. The phantom dominated late time solution appears in the context of quintom cosmologies with exponential potentials when one resorts to standard dynamical systems techniques as it has been shown in \cite{Zhang:2005eg,Guo:2004fq} and \cite{Lazkoz:2006pa}. The quintom is a hybrid of a quintessence component, usually modeled by a real scalar field that is minimally coupled to gravity, and a phantom field: another real scalar field --minimally coupled to gravity-- with negative kinetic energy \cite{Wei:2005fq,Wei:2005nw,Feng:2004ad,Feng:2004ff, Xia:2004rw, Zhao:2005vj,Wu:2005ap,Zhang:2005kj,Cai,Guo:2004fq,Zhang:2005eg,Wei:2005si,Lazkoz:2006pa,MohseniSadjadi:2006hb, Alimohammadi:2006tw}.\footnote{Quintom-like behavior (with $w=-1$ crossing) has been found also in the context of holographic dark energy \cite{holographic1,holographic2,holographic3}.} 

The aim of the present paper is, precisely, to provide further elements to support the interest on quintom DE models. We do so by studying from the dynamical systems perspective quintom cosmologies with arbitrary potentials (other than exponential ones). In this regard, our study is complementary, in some sense, to that of references \cite{Zhang:2005eg, Guo:2004fq, Lazkoz:2006pa}. As customary, this kind of analysis relies on an appropriate choice of phase-space coordinates, which builts partially on those made in precedent references \cite{Lazkoz:2006pa, Nunes:2000yc}.

Perhaps, the most important conclusion of the present research is confirming the result of reference \cite{Lazkoz:2006pa} (the existence of phantom attractors is not generic) in the more general setup of quintom DE with arbitrary potentials (other than exponential ones).\footnote{The same conclusion was reached  in reference \cite{Wei:2005fq}, where a hessence model with exponential and inverse power law potentials were studied from the dynamical systems view point.} Peculiarly, in our framework, de Sitter ($w=-1$) attractors can be associated only with saddle points of (the natural logarithm of) the potential.

The paper has been organized as follows. In section II we give the details of the dark energy model under investigation. In section III we turn to perform the dynamical study of this quite general cosmological model for finite values of the component scalar fields. For illustration purposes, we choose a class of potentials which was derived in  the literature on inflation. Then, and in order to complement the dynamical system analysis carried out in section III, we make in section IV a different choice of phase space variables relating the scalar fields, which allows to study the asymptotic regime where both component scalar fields diverge. Conclusions are given in section V. Finally, to make the paper self-contained, we have added an Appendix with some recipes on dynamical systems and its applications. Throughout this paper we use natural units.

\section{The model}

We will investigate the evolution of a spatially flat Friedmann-Robertson-Walker (FRW) spacetime fuelled by dark matter (dust) with energy density $\rho_m$, and by quintom dark energy with energy density $\rho_{de}$ and pressure $p_{de}$ given respectively by

\ben
\rho_{de}=\frac{1}{2}\dot\phi^2-\frac{1}{2}\dot\varphi^2+V(\phi,\varphi),\,
 p_{de}=\omega \rho_{de}, \een
where for our quintom model, the scalar field $\varphi$ plays the role of a phantom field. In the above equations
\be w=\frac{\dot\phi^2-\dot\varphi^2-2V(\phi,\varphi)}{\dot\phi^2-\dot\varphi^2+2V(\phi,\varphi)}\ee is the equation of state (EOS) parameter of the dark energy. Thus, we are assuming our dark energy comprises the contribution of
two  scalar fields $\phi$ and $\varphi$ interacting through their potential energy $V(\phi,\varphi)$, and satisfying separate energy conservation equations, namely:
\ben
&\ddot\phi+3H\dot\phi+\partial_{\phi}V(\phi,\varphi)=0 \label{consphi},\\
&\ddot\varphi+3H\dot\varphi-\partial_{\varphi}V(\phi,\varphi)=0 \label{consvarphi}.
\een

We adhere here to Einstein's theory of general relativity so, as often done, dark energy will be assumed to be minimally coupled to matter.
With our assumptions, the cosmological equations are the standard Friedmann equation
\be
3 H^2=\rho_m+\rho_{de},\label{F}\\
\ee
 the continuity equation
\ben
&&\dot\rho_m+3H\rho_m=0\,\label{consm},
\een
and the conservation equations Eqs. (\ref{consphi}, \ref{consvarphi}).
Here and throughout overdots denote differentiation with respect to cosmic time $t$, $H=\dot a/a$ is the Hubble factor, and $a$ is the scale factor.

Combining Eqs. (\ref{consphi}-\ref{consm}) one obtains the evolution equation for $\dot H$:
\begin{equation}
-2 \dot H =\rho_m+\rho_{de}+p_{de}.\label{ray}
\end{equation} Equations (\ref{consphi}-\ref{ray}) are the basic set required for formulating a dynamical systems investigation of the evolution of these cosmological models.

\section{Phase-space}

It is well-known that information about the evolution of cosmological models can be retrieved using dynamical systems as a tool. Particularly, the asymptotic (and sometimes the intermediate) behavior of cosmological models are closely related to concepts like past and future attractors \cite{reza,Wainwright:2004cd}.
Our purpose, thus, is to take advantage of that fact, and apply it to the model under consideration.

We want to construct a dynamical system with the following properties:
\begin{enumerate}
\item It is autonomous.
\item It has the lowest possible dimensionality.
\item It describes a physically interesting and reasonable configuration within the framework of quintom cosmologies.
\end{enumerate}

In order to construct a dynamical system with these properties one
of the first steps is to introduce a set of convenient (in most
occasions expansion normalized \cite{Wainwright:2004cd}) variables
which allow  rewriting the conservation equations and the
evolution equation of $H$ as a dynamical system subject to a
constraint arising from the Friedmann equation (\ref{F}). Since we
want to study arbitrary potentials (as arbitrary as possible, but
with the exclusion of exponential like potentials),  a minimum of
six variables is needed to construct an autonomous dynamical
system. However, the problem gets somewhat simplified --at least
concerning its dimensionality-- if we use the constraint arising
from Eq. (\ref{F}) as a definition for one of the variables.
Hence, the minimal number of dimensions of our model is lowered down
to five.

The next step in the study of the evolution of our dark energy model viewed as a  dynamical system is to find its fixed (or critical) points. The stability of the fixed points is then analyzed by studying the  linearized dynamical system obtained by expanding the evolution equations about those fixed points (see \cite{hirsch,reza} for seminal references and the Appendix for useful recipes).

Turning back to the first step, it follows that one can present Eqs. (\ref{consphi}-\ref{ray}) in the form of a dynamical system by making the following choice of variables: $(x_\phi,\,x_\varphi,\,y,z,\,\phi,\varphi)$, where
\begin{eqnarray}
&&x_\phi=\frac{\dot\phi}{\sqrt{6}H},\;x_\vphi=\frac{\dot\vphi}{\sqrt{6}H},\;\\ &&y=\frac{\sqrt V}{\sqrt 3 H},\;z=\frac{\sqrt {\rho_m}}{\sqrt 3 H},\label{vars}
\end{eqnarray} and $\phi$, $\varphi$ are the scalar field variables. This choice of phase-space variables renders the Friedmann equation as
\begin{eqnarray}
x_\phi^2- x_\vphi^2+y^2+z^2=1.
\label{ct}
\end{eqnarray}

The constraint (\ref{ct}) follows from Eq. (\ref{F}) and it allows, as we have noticed before, to consider only the evolution of those five variables other than $z$ (the evolution of which is being determined by the evolution of the former ones).

In what follows we will restrict ourselves to $y\ge0$ and to $H\ge0$ since the main concern of this paper are  expanding universes.\footnotemark\footnotetext{Notice that the equations (\ref{eqxphi}-\ref{eqvarphi}) are invariant under the change $y\rightarrow -y$.} Combining expressions (\ref{consphi}-\ref{vars}), the following evolution equations are obtained:

\ben
&&x_\phi'=\frac{1}{3} \left(-\frac{\sqrt{6}}{2} y^2 \partial_\phi \ln V+(q-2) x_\phi\right),\label{eqxphi}\\
&&x_\vphi'=\frac{1}{3} \left(\frac{\sqrt{6}}{2} y^2\partial_\varphi \ln V+(q-2) x_\varphi \right),\label{eqxvarphi}\\
&&y'=\frac{1}{3} (1+q-\frac{\sqrt{6}}{2}(x_\phi\partial_\phi \ln V+x_\varphi\partial_\varphi \ln V)) y,\label{eqy}\\
&&\phi'=\frac{\sqrt{6}}{3}x_\phi,\label{eqphi}\\
&&\varphi'=\frac{\sqrt{6}}{3}x_\varphi.\label{eqvarphi}\een Here primes denote differentiation with respect to a new time variable $\tau=\log a^3$, and $q\equiv-\ddot a a/\dot a^2$ stands for the deceleration factor. Explicitly,
\be
q=\frac{1}{2} \left(3 \left(x_\phi^2- x_\varphi^2-y^2\right)+1\right).
\ee

The evolution equations of variables $x_{\phi}$, $x_{\vphi}$, $y$, $\phi$ and $\varphi$ form a 5D dynamical system  defined on the  phase space

\ben\Psi=\{(x_\phi,x_\vphi,y):0\le x_\phi^2- x_\vphi^2+y^2\le1\}\times\nonumber\\\{(\phi,\varphi)\in\mathbb{R}^2\}.\een

The phase space $\Psi$ is unbounded because the variables $x_\varphi,\,\phi,\,\varphi$ can reach infinite values. However the projection of the phase space into the subspace $(x_\phi,\,x_\varphi,\,y)$ is limited by the hypersurface defining the hyperboloid.

In general, the phase portrait can be split into two regions depending on the values of $\phi$ and $\varphi$. One region is defined by the condition that both scalar fields are finite. This region comprises matter dominated and de Sitter solutions. In the second region both scalar fields diverge.
As we shall see, it is the latter region the one that may accommodate phantom solutions whereas the former cannot, that, is, phantom behavior only appears in the regime where the scalar fields blow up.
In this regard we have to notice that the above choice of phase-space variables (\ref{eqxphi}-\ref{eqvarphi}) is not adequate to study the region where scalar fields diverge, so a different choice of variables is necessary. This case (infinite ($\phi,\vphi$)) will be studied separately in section IV.

\subsection{Matter dominated solutions}

The dynamical system (\ref{eqxphi}-\ref{eqvarphi}) admits the biparametric class of non-hyperbolic critical points  $(x_\phi,x_\vphi,y)=(0,0,0)$, for arbitrary $(\phi,\varphi)\in\mathbb{R}^2$. These critical points, which we will
denote generically as $O$, represent matter dominated cosmological solutions. Typically, the scale factor and the matter energy density evolve as $a\propto t^{2/3}$ and $\rho_m\propto t^{-2}$ respectively. This class of solutions is relevant from the end of the epoch of radiation domination till the very recent epoch when the dark energy contribution equated the contribution from the dark matter.

Making a local linear analysis we obtain that the eigenvalues of the linearization around these critical points (with $\phi$ and $\varphi$ fixed) are $(-{1}/{2},-{1}/{2},{1}/{2},0,0).$ Due to the existence of two null eigenvalues, as already said, these critical points are non-hyperbolic so that the linear analysis is not conclusive in this case. Alternatively, one can determine analytically the stable, unstable and center manifolds of those critical points (see the Appendices A and B).

\subsection{de Sitter solutions}

The dynamical system (\ref{eqxphi}-\ref{eqvarphi}) admits solutions where  $(x_\phi,x_\vphi,y)=(0,0, 1)$, for each $(\phi^\star,\varphi^\star)\in\mathbb{R}^2$ such that $\partial_\phi \ln V(\phi^\star,\varphi^\star)=\partial_\varphi \ln V(\phi^\star,\varphi^\star)=0,$ that is, $(\phi^\star,\varphi^\star)$ are the stationary points (extrema or saddle points) of $\ln V$.\footnotemark\footnotetext{The stationary points of $\ln V$ coincide in general with the stationary points of $V$ provided these are not zeros of the potential.}

For fixed $\phi^\star$ and $\varphi^\star$ the corresponding
solution is dominated by the potential energy of the quintom
field, which means it is a de Sitter solution, so we will denote
this class of fixed points as $dS.$ The scale factor evolves
according to the following exponential law $ a\propto
\exp\left[\sqrt{V(\phi^\star,\varphi^\star)/3}\, t\,\right].$
These solutions are expected to be important at late times when the potential energy mimics an effective
cosmological constant.

The eigenvalues of the linearization around the corresponding fixed points are
\ben
&& \lambda_1=-1,\\
&&\lambda_{2}^{\pm}=-\frac{1}{2}\pm\frac{1}{2}\sqrt{1+\frac{2}{3}\left({\Delta_1-\sqrt{\Delta_2}}\right)},\\
&&\lambda_{3}^{\pm}=-\frac{1}{2}\pm\frac{1}{2}\sqrt{1+\frac{2}{3}\left({\Delta_1+\sqrt{\Delta_2}}\right)}
\een where

\be\Delta_1=\left[\frac{\partial^2 \ln V}{\partial\varphi^2}-\frac{\partial^2 \ln V}{\partial
\phi^2}\right]_{\phi=\phi^*, \varphi=\varphi^*},\label{D1}\ee

\ben\Delta_2=[\left(\frac{\partial^2 \ln V}{\partial\phi^2}+ \frac{\partial^2 \ln V}{\partial
\varphi^2} \right)^2\nonumber\\-4\left(\frac{\partial^2 \ln V}{\partial
\varphi\partial\phi}\right)^2]_{\phi=\phi^*,\varphi=\varphi^*}.\label{D2}\een

If the point $(\phi^\star,\varphi^\star)$ is an extremum of $\ln V$ then the quantity $\Delta$ defined by

\be\Delta=\left[\frac{\partial^2 \ln V}{\partial \varphi^2}\frac{\partial^2 \ln V}{\partial
\phi^2}-\left(\frac{\partial^2 \ln V}{\partial\varphi\partial\phi}\right)^2\right]_{\phi=\phi^*, \varphi=\varphi^*}\label{delta}\ee is positive; if it is a minimum (maximum) then ${\partial^2\ln V}/{\partial \phi^2}> 0 \;
(< 0)$ and ${\partial^2\ln V}/{\partial \varphi^2}> 0 \;(< 0).$ If $\Delta< 0$, then $(\phi^\star,\varphi^\star)$ is a saddle point of $\ln V$.

From the former analysis it follows that the dynamical behavior of
the critical points in $\Psi$ depends on the classification of the
critical points of $\ln V$: $(\phi^\star,\varphi^\star)$.

\subsubsection{Dynamical behavior of the critical points in $\Psi$ associated with the de Sitter phase}

If $(\phi^*, \varphi^*)$ is an extremum of $\ln V(\phi,\varphi),$  then the condition $\Delta> 0$ implies $\Delta_2> \Delta_1^2.$ In this case the corresponding critical point in $\Psi$ is a saddle; i.e., all the real parts of the eigenvalues of the matrix {\bf A} evaluated at the corresponding critical point are different from zero and at least two of them have different signs (see the Appendices A and B).

If $(\phi^*, \varphi^*)$ is a saddle point of $\ln V(\phi,\varphi),$ then the condition $\Delta< 0$ implies $\Delta_2< \Delta_1^2.$ In this case the corresponding critical point in the phase space $\Psi$ is an attractor in the following three cases\footnotemark\footnotetext{The critical point is an the attractor if all the real parts of the eigenvalues of the matrix {\bf A} evaluated at the corresponding critical point are negative (see the Appendices A and B).}:

\begin{itemize}
\item Case i) All the eigenvalues are negative reals. This is the case if 
$$\qquad 0\leq\Delta_2<{9}/{16} \mbox{ and }-{3}/{2}+\sqrt{\Delta_2}< \Delta_1 <-\sqrt{\Delta_2}.$$
\item Case ii) $\lambda_2^\pm$ are complex conjugate eigenvalues and $\lambda_3^\pm$ are negative real eigenvalues. This is the case if $$\qquad0< \Delta_2< {9}/{16}\mbox{ and }-\sqrt{\Delta_2}< {3}/{2}+\Delta_1 \leq \sqrt{\Delta_2},$$ or $$\qquad\Delta_2\geq{9}/{16} \mbox{ and } -{3}/{2}-\sqrt{\Delta_2}< \Delta_1 <-\sqrt{\Delta_2}.$$
\item Case iii) $\lambda_2^\pm$ and $\lambda_3^\pm$ are respectively complex conjugate eigenvalues. This is the case if $$\Delta_2\geq 0\mbox{ and }\Delta_1\leq -{3}/{2}-\sqrt{\Delta_2}.$$
\end{itemize}

\begin{figure}
\begin{center}
\hspace{0.3cm}
\includegraphics[width=6.5cm,height=5cm]{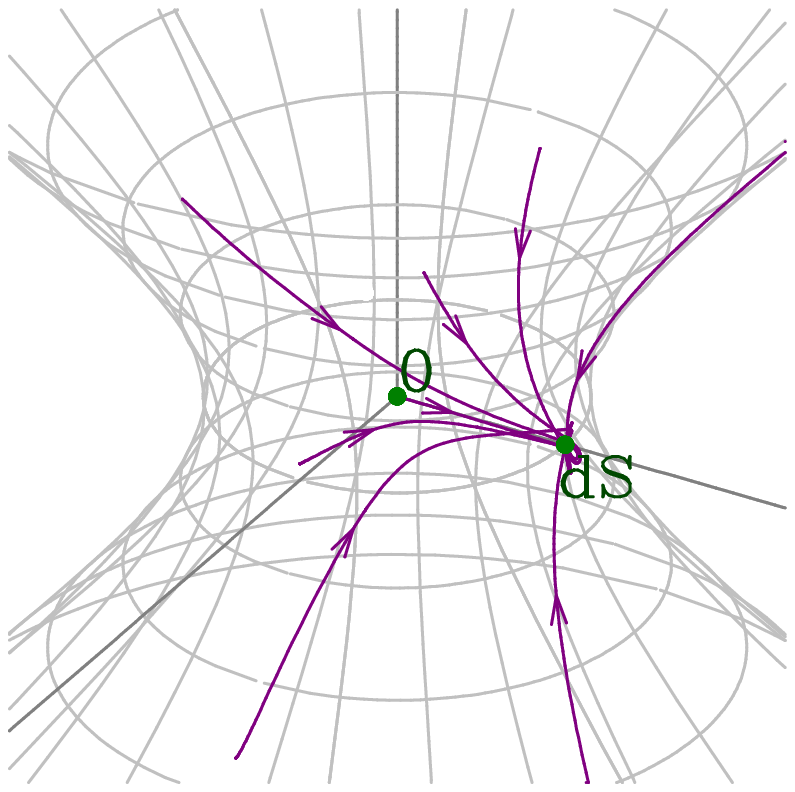}
\put(-180,5){$x_{\phi}$}
\put(-90,130){$x_{\varphi}$}
\put(-10,58){$y$}
\vspace{0.3cm}
\begin{center}(a)\end{center}
\vspace{0.3cm}
\includegraphics[width=6.5cm,height=5cm]{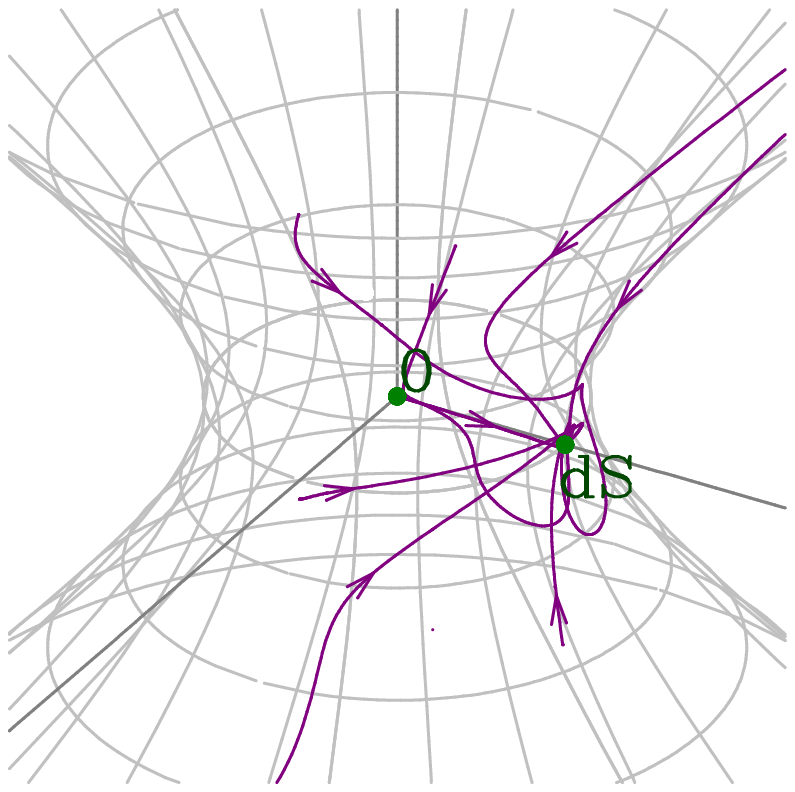}
\put(-180,5){$x_{\phi}$}
\put(-90,130){$x_{\varphi}$}
\put(-10,58){$y$}
\vspace{0.3cm}
\begin{center}(b)\end{center}
\vspace{0.3cm}
\includegraphics[width=6.5cm,height=5cm]{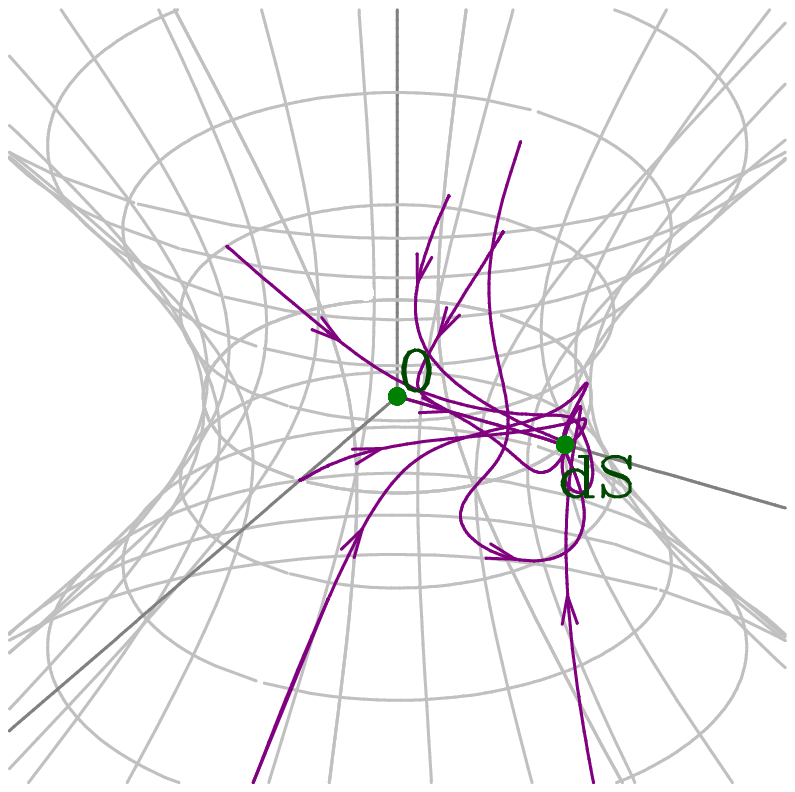}
\put(-180,5){$x_{\phi}$}
\put(-90,130){$x_{\varphi}$}
\put(-10,58){$y$}
\vspace{0.3cm}
\begin{center}(c)\end{center}
\vspace{0.3cm}
\caption{Projections of the phase-space trajectories for different values of $M,$ $m,$ $g,$ and $\lambda$:
 $M=4,\,m=3,\,g=2,$ $\lambda=2$ (a);
 $M=2,\,m=1.5,\,g=2,$ $\lambda=1.3$ (b) and
 $M=4,\,m=5,\,g=2,$ and $\lambda=3.5$ (c).
The point $dS$ represents the de Sitter attractor. All the orbits portrayed have been truncated so that $y\ge0$ and they have been obtained by fixing the initial values for the scalar fields to $\phi(\tau=0)=0.20$ and $\varphi(\tau=0)=0.19.$}
\label{phase}
\end{center}
\end{figure}


\subsection{Example: hybrid inflation potential}

To illustrate the former result we will consider a two scalar fields model with potential

\be V(\phi,\varphi)=\frac{1}{2}m^2\phi^2+\frac{1}{2} g^2\phi^2\varphi^2+\frac{(M^2-\lambda\,\varphi^2)^2}{4\lambda},\label{hybridpotential}\ee
where $m,\;g,\;M,\;\lambda$ are real-valued non-negative constants.\footnote{Quintom cosmologies with polynomial potentials were investigated in \cite{Arefeva,sergey}, where only dark energy (no CDM) was considered.}

The potential (\ref{hybridpotential}) has been formerly considered in \cite{Linde:1991km,Kofman:1986wm,Kofman:1988xg,Salopek:1988qh,Kofman:1991qx,Hodges:1989dw,Linde:1993cn} in the context of hybrid inflation. Since our context is late universe, the virtues of this potential regarding (early time) inflation are not of interest to us, we have selected it because its functional form is suited to illustrate some of the peculiar
features of the late time evolution of our quintom dark energy models.

It is straightforward to show that $\ln V(\phi,\varphi)$ has a unique real-valued stationary point at $(\phi^\star,\varphi^\star)=(0,0)$ which is a saddle point of $\ln V$ since $\Delta(0,0)=-{16\lambda^2 m^2}/{M^6}<0$ (see eq. (\ref{delta})). This means that the de Sitter phase is an attractor in $\Psi$. Actually, since the function $\ln V$ with $V$ given by (\ref{hybridpotential}) has only a saddle point (it has no minimum at all), for a quintom model the critical point (in the phase space $\Psi$) associated with $(\phi^\star,\varphi^\star)$ is stable in the following three cases as follows from the former analysis (see cases i), ii), and iii) in the former subsection):

\begin{itemize}
\item Case i) This will the case (see Fig.  \ref{phase}a) if either
$$0<M<m\mbox{ and }0<\lambda<{3 M^4}/{16\, m^2}$$
or if 
$$0<m\leq M\mbox{ and }0<\lambda<{3 M^2}/{16}.$$ 
\item Case ii) This case will occur (see Fig. \ref{phase}b)  if either 
$$0<M<m\mbox{ and }{3 M^4}/{16 m^2}\leq\lambda<{3 M^2}/{16}$$or if 
$$0<m\leq M\mbox{ and }{3 M^2}/{16}\leq\lambda<{3 M^4}/{16\, m^2}.$$ 
\item Case iii) Finally, one will have this case (see Fig. \ref{phase}c) if either $$0<M<m \mbox{  and }\lambda\geq{3 M^2}/{16}$$
 or if $$0<m\leq M\mbox{ and }\lambda\geq{3 M^4}/{16\, m^2}.$$ 
\end{itemize}

This completes the study of the region with no divergent scalar fields.

\section{Equilibrium points at infinite $(\phi,\,\varphi)$}

One of the most widespread view points within the framework of (DE) scalar field cosmologies is that only exponential potentials may be responsible for a tracking phase. For instance, exponential potentials have been considered as a source of interaction between scalar fields in conventional \cite{Copeland:1999cs, Coley:1999mj, Aguirregabiria:2000hx, Blais:2004vt} and unconventional \cite{Guo:2004fq, Zhang:2005eg, Lazkoz:2006pa} cosmologies. However, in reference \cite{Wei:2005fq}, it was shown that in the context of quintom-like DE cosmologies (hessence models) with exponential and (inverse) power law potentials, some stable attractors can exist which are either scaling solutions or hessence-dominated solutions with EOS larger than or equal to -1. Similar results will be confirmed in what follows.

In \cite{Lazkoz:2006pa} a step forward was done with respect to \cite{Guo:2004fq, Zhang:2005eg}: The potential used in the former reference
\begin{equation}V=V_0
e^{-\sqrt{6}\left(\tilde{n}\phi+\tilde{m}\varphi\right)},\label{expot}\end{equation}
leads to the existence of an scaling regime; i.e., \emph{it was proved that quintom cosmologies with exponential potentials were not devoid of asymptotic scaling phases}.
In the context of this study, the question that comes to mind is the following: \emph{can there exist tracking solutions in the quintom scenario with arbitrary potentials?}
This question is partially motivated by the fact that  tracking behavior  may exist in single scalar fields models with potentials that asymptotically tend to exponential ones \cite{Nunes:2000yc}, so it might well be the case that in our context
a broad family of potentials satisfying some particular requirement regarding their asymptotic functional form could display tracking behavior.

The existence of tracking phases in the region of finite $\phi$ and $\varphi$ is precluded by our previous analysis. So one has to turn attention to the region where
 $\phi=\pm\infty$ and $\varphi=\pm\infty$. Thus, our purpose is to try and investigate the possible existence of scaling phases in quintom cosmologies with arbitrary (other than the exponential) potentials, by concentrating on the region where the scalars fields diverge. This is to be done following the method developed in \cite{Nunes:2000yc}. It must be stressed that our study of this region will be done in full detail, and even though emphasis will be given to tracking solutions, other possible asymptotic states will be
reported in case they exist.  This analysis will be complementary to the dynamical study of section III.

\begin{table*}[t!]\caption[crit]{Location, existence and deceleration factor of the critical points for $\tilde{m}>0$,  $\tilde{n}>0$ and $y>0$, in the infinite $(\phi,\varphi)$ region.}
\begin{tabular}{@{\hspace{4pt}}c@{\hspace{14pt}}c@{\hspace{14pt}}c@{\hspace{14pt}}c@{\hspace{18pt}}c@{\hspace{18pt}}c@{\hspace{2pt}}}
\hline
\hline\\[-0.3cm]
Name &$x_\phi$&$x_\vphi$&$y$&Existence&$q$\\[0.1cm]
\hline\\[-0.2cm]
$O$& $0$& $0$& $0$& All $\tilde{m}$ and $\tilde{n}$ &$\ds\frac{1}{2}$\\[0.2cm]
$C_{\pm}$ & $\pm\sqrt{1+{x_{\varphi}^*}^2}$& $x_{\varphi}^*$& $0$&
All $\tilde{m}$ and $\tilde{n}$ &$2$ \\[0.2cm]
${P}$&
$\tilde{m}$& $-\tilde{n}$ &$\sqrt{1-\tilde{m}^2+\tilde{n}^2}$&$\tilde{m}^2-\tilde{n}^2< 1$&$-1+3(\tilde{m}^2-\tilde{n}^2)$\\[0.2cm]
${T}$& $\ds\frac{\ds\tilde{m}}{\ds2(\tilde{m}^2-\tilde{n}^2)}$&
$-\ds\frac{\ds\tilde{n}}{\ds2(\tilde{m}^2-\tilde{n}^2)}$&
$\ds\frac{1}{2\sqrt{\tilde{m}^2-\tilde{n}^2}}$&$\tilde{m}^2-\tilde{n}^2\ge1/2$&$\ds\frac{1}{2}$\\[0.4cm]
\hline \hline
\end{tabular}\label{tab1}
\end{table*}

We start by making a different choice of phase space variables relating the scalar fields: $u=\phi^{-1},$
$v=\varphi^{-1}.$ The latter variables are complemented by $x_\phi,\,x_\vphi$, and $y$ defined in Eqs.
(\ref{vars}). Making use of the required coordinate transformations the original system (\ref{eqxphi}-\ref{eqvarphi}) can be expressed as
\begin{eqnarray}
&& x_\phi'=\frac{1}{3}\left((-2+q)x_\phi+3\tilde{N}y^2\right)
,\label{eqxphimod}\\
&&
x_\varphi'=\frac{1}{3}\left((-2+q)x_\varphi-3
\tilde{M} y^2\right),\label{eqxvarphimod}\\
&& y'=\frac{1}{3}y\left(1+q-3\left(\tilde{N}x_\phi
+\tilde{M}x_\varphi \right)\right),\label{eqymod}\\
&&u'= -\frac{\sqrt{6}}{3}u^2 x_\phi\\
&&v'= -\frac{\sqrt{6}}{3}v^2 x_\varphi
\end{eqnarray}
where $\tilde{V},$ $\tilde{N}$ and $\tilde{M}$ are arbitrary functions of $u$ and $v$ defined, respectively by:
\begin{eqnarray}
&&\tilde{V}(u,v)=V(u^{-1},v^{-1}),\\
&&\sqrt{6}\left(\tilde{N},\tilde{M}\right)=\left(
u^2\frac{\partial\ln \tilde{V}}{\partial
u},v^2\frac{\partial\ln\tilde{V}}{\partial
v}\right).
\end{eqnarray} 
We assume $\tilde{V}$ is such that
\begin{equation}\lim_{(u,v)\rightarrow (0,0)} \left(
u^2\frac{\partial\ln \tilde{V}}{\partial
u},v^2\frac{\partial\ln\tilde{V}}{\partial
v}\right)=\sqrt{6}(\tilde{n},\tilde{m})\label{exponential}.
\end{equation}

The requirement (\ref{exponential}) implies that the only essential feature of the potential is that it should behave asymptotically as an exponential potential. This (very) general condition on the potential $V$, leads us to study the phase-space trajectories in the invariant subset $(u,v)=(0,0)$ by taking the limit $(u,v)\rightarrow (0,0)$ in the equations  (\ref{eqxphimod}-\ref{eqymod}). By continuity, the reduced dynamical system in that invariant subset is
\begin{eqnarray}
&&x_\phi'=\frac{1}{3} \left(3 \tilde{m} y^2+(q-2) x_\phi\right),\label{eqxphiasymp}\\
&&x_\vphi'=-\frac{1}{3} \left(3 \tilde{n} y^2-(q-2) x_\varphi \right),\label{eqxvphiasymp}\\
&&y'=\frac{1}{3} (1+q-3(\tilde{m} x_\phi+\tilde{n} x_\varphi)) y\label{eqyasymp}
\end{eqnarray}

The equations (\ref{eqxphiasymp}-\ref{eqyasymp}) are the same as in \cite{Lazkoz:2006pa}, but the subtle point here is that the dynamical system (\ref{eqxphiasymp}-\ref{eqyasymp}) describes the dynamical behavior of a quintom cosmology with an arbitrary (asymptotically exponential) potential in an invariant subset of $\mathbb{R}^5$. So the dynamics is richer than in \cite{Lazkoz:2006pa}. Nevertheless, the results reported in the latter reference apply to our more general case. These are immediately summarized.

As we have noticed before, the first approach to analyzing the evolutionary behavior of a cosmological model from dynamical systems point of view is to find the critical points of the system arising when normalized
variables are taken into account. Particularly at the invariant subset $(u,v)=(0,0)$ there may exist representatives of the critical points $O,\,C_\pm,\,P$ and $T$ studied in
\cite{Lazkoz:2006pa}.\footnotemark\footnotetext{For fixed points of class $O$ the interpretation in Section IIIA is
applicable as well.} In Table \ref{tab1} the location, the existence and deceleration factor of these critical points for $\tilde{m}>0$,  $\tilde{n}>0$ and $y>0$ are presented. 

The existence of a scaling regime of is represented by the critical point  $T$ for values of $\tilde{m}$ and
$\tilde{n}$ such that $\tilde{m}^2-\tilde{n}^2\geq 1/2.$ Whenever it exists, it
represents a attractor solution in which quintom dark energy tracks matter
(the equation of state of the quintom fluid is dust-like; i. e.,
$w=0$). This late-time asymptotic state represents a decelerating cosmological
model in which the fractional energy densities of matter and dark
energy are proportional. Particularly, $\Omega_{de}$ and $\Omega_m$ are given by
$\Omega_{de}=(2\tilde{m}^2-2\tilde{n}^2)^{-1},$ and
$\Omega_m/\Omega_{de}=-1+2\tilde{m}^2-2\tilde{n}^2\geq 0.$ The dynamical character of $T$ depends on the value of $\Omega_{de}$: it is either an stable focus or an stable node if $\Omega_{de}<7/8,$ or $7/8\leq \Omega_{de}<1$, respectively. 

The fixed point $P$ represents a solution in which quintom dark energy dominates over matter (the equation of state of the quintom fluid corresponds to a fluid that redshifts faster than dust). When such point exists it is an attractor only when its existence precludes that of the point T. This late-time asymptotic state does not necessarily represents an accelerating cosmological model, that depends on the quantity $\tilde{m}^2-\tilde{n}^2$. Given its eigenvalues structure (see table \ref{tab1}), this solution is either a saddle or a stable node. The accelerated solutions associated with this fixed point can provide a good representation of the presently observed universe. 

It is noticeable that in the regime ($\phi\rightarrow\infty,\;\varphi\rightarrow\infty$) phantom solutions may exist as well. Actually, a potential satisfying (\ref{exponential}) allows $P$ to be characterized by $w<-1$ (i.e. phantom dark energy) if $\tilde{m}^2<\tilde{n}^2$. Other possibilities include the cases $w=-1$ (i.e. de Sitter vacuum dark energy) if $\tilde{m}=\tilde{n}$, and $-1<w<-1/3$ (i.e. quintessence dark energy) if $0<\tilde{m}^2-\tilde{n}^2<1/3.$ These are the solutions which the interest on quintom cosmologies is grounded on.  

In general the potential $\tilde{V}$ responsible for such a scaling regime can be reconstructed from functions $\tilde{N}$ and $\tilde{M}$  satisfying:

$${v^{-2}}\frac{\partial\tilde{M}}{\partial u}={u^{-2}}\frac{\partial \tilde{N}}{\partial v},\;  \lim_{(u,v)\rightarrow (0,0)}(\tilde{N},\tilde{M})=(\tilde{n},\tilde{m})$$  as
\begin{eqnarray} \tilde{V}(u,v)=V_0\exp\int_1^u \frac{\sqrt{6} \tilde{N}(\mu ,v)}{\mu ^2} \, d\mu\times\nonumber\\\exp\int_1^v \left(\frac{\sqrt{6} \tilde{M}(u,\nu )}{\nu ^2}-\int_1^u \frac{\sqrt{6} \partial_{\nu}\tilde{N}(\mu ,\nu )}{\mu ^2} \, d\mu \right)\, d\nu.\end{eqnarray}
Note that $$\frac{\partial^2\ln\tilde{V}}{\partial u\partial
v}=\frac{\sqrt{6}}{v^2}\frac{\partial \tilde{M}}{\partial u}.$$

Finally, this region of the space phase admits non-isolated fixed points which build the hyperbolae $C_{\pm}$ reported in Table I. Given that their interpretation involves technicalities which do not add much to the current discussion we submit the reader to \cite{Lazkoz:2006pa} for further details.

Summarizing, in this subsection we have reinforce our results in \cite{Lazkoz:2006pa}, as we have shown that tracking solutions can exist in quintom cosmologies. We have identified that (provided they exist) these tracking solutions can be found in the region $\phi=\pm\infty$ and $\varphi=\pm\infty$.\footnotemark\footnotetext{This feature was first noticed in \cite{Nunes:2000yc} but in the framework of single scalar fields.} In addition, we have shown that the region $\phi=\pm\infty$ is the only possible locus for the phantom cases.

\section{Conclusion}

The dynamical systems analysis is a very powerful tool when one wants to extract useful information about the dynamics of the cosmological evolution without knowledge of exact solutions and of specific initial conditions.

In the present paper we have applied dynamical systems tools to retrieve information about the dynamics of quintom models with arbitrary potentials (other than the exponential ones). Appropriate phase space variables has been chosen so as to allow for a study of regions in the ($\phi$,$\vphi$)-plane for both finite and infinite values of the scalar field variables $\phi$, $\vphi$.

The results of the present study indicate that existence of phantom attractors is not generic; it depends on the features of (the natural logarithm of) the potential. For quintom models, in the finite ($\phi$, $\vphi$)-region, there may exist either de Sitter attractors associated with the saddle points of $\ln V$, or tracking attractors in the infinite ($\phi$, $\vphi$)-region. For the example studied in section III C (potential (\ref{hybridpotential})) the de Sitter phase represents an attractor solution of the phase-space $\Psi.$

We want to underline that the present study is not particularly well suited for potentials which are products of sums  of exponentials on each of the scalar fields. The reason is that in the latter cases dimensional reduction of the dynamical system is possible (and performing it is strong advisable). Therefore, our study should be regarded as   complementary to that of references \cite{Zhang:2005eg, Guo:2004fq, Lazkoz:2006pa}.

\acknowledgments R.L. is supported by the University of the Basque Country through research grant UPV00172.310--14456/2002 and by the Spanish Ministry of Education and Culture through the RyC program, and research grants FIS2004--01626 and FIS2005--01181. G.L. and I.Q. acknowledge the MES of Cuba by partial financial
support of the present research. G. L. wishes to thank also the hospitality of the Dept. of Theoretical Physics of the University of the Basque Country during the completion of part of this work.

\appendix

\section{Dynamical systems recipes}

The first step to obtain qualitative information about the solutions of a differential equation
$\mathbf{x}'=\mathbf{f}(\mathbf{x})$ is to study the flow of the differential equation in the vicinity of the critical points, i.e, to study their stability \cite{reza}. The essential idea is to linearize the differential equation at each fixed point and then use the Hartman-Grobman theorem which in colloquial terms can be stated as follows: \emph{the flow of a non linear differential equation in the neighborhood of a non--hyperbolic critical point can be deformed continuously into the flow of its linearization}; i.e, the orbits of both systems are (locally) qualitatively the
same.

Since the flow of a linear differential equation can be obtained explicitly, we will present the analysis for linear systems.

For the linear differential equation $\mathbf{x}'=\mathbf{A}\mathbf{x}$ defined in
$\mathbb{R}^n$ (or defined in a subset of the whole real space; or in general defined in a smooth manifold) we can determine the eigenvalues of the matrix $\mathbf{A}$ (complex in general and not necessarily different)
and the associated eigenvectors which span three subspaces of $\mathbb{R}^n$ (which depend on the nature of the associated eigenvalues): $E^s$, $E^u$ y $E^c.$ The former subspaces contain disjoint orbits which form a partition of the state space, i.e. $E^s\otimes E^u\otimes E^c=\mathbb{R}^n$. They are, respectively, the stable subspace (spanned by the eigenvectors whose associated eigenvalues has negative real parts), the unstable subspace (spanned by the eigenvectors whose associated eigenvalues has positive real parts), and the central subspaces which are generated by the eigenvectors whose associated eigenvalues have zero real parts. 

The stable and unstable subspaces are characterized, respectively, by the properties: 
\ben\mathbf{x}\in E^s\Rightarrow\lim_{\tau\rightarrow\infty}e^{\mathbf{A}\tau}\mathbf{x}=0,\nonumber\\
\mathbf{x}\in E^u\Rightarrow\lim_{\tau\rightarrow -\infty} e^{\mathbf{A}\tau}\mathbf{x}=0.\nonumber\een 
These describe the asymptotic behavior: all the initial states in the stable subspace are attracted by the critical point $\mathbf{x}=0$ and all the initial states in the unstable subspace are repelled by $\mathbf{x}=0.$

If the system $\mathbf{x}'=\mathbf{f}(\mathbf{x})$ is non linear, we can use the Hartman-Grobman theorem.
Using it we can define the manifolds $\mathcal{E}^{(s,u,c)}$ (stable, unstable and center manifold respectively) at a fixed point, those manifolds are tangent to the corresponding subspaces $E^{(s,u,c)}$ of the linearization at the fixed point (stable, unstable and center subspaces respectively). All the orbits in $\mathcal{E}^s$ converges  asymptotically to the fixed point when the time variable evolves ($\tau\rightarrow\infty$), whereas all the orbits at $\mathcal{E}^u$  converge asymptotically to the fixed point when $\tau\rightarrow -\infty.$ The manifold $\mathcal{E}^c$ contains all the orbits whose asymptotic behavior can not be analyzed making use of the linear analysis.

\section{Applications}

Our dynamical systems were constructed making use of a maximum of five dynamical variables $x^i$, $i=1\dots 5$. The corresponding phase-space equations can be written symbolically by  
\be {x^i}\hspace{1pt}'=f^i(x^1,x^2,\dots,x^5),\ee where prime denotes the derivative with respect to the alternative time variable $\tau$ which is chosen by convenience.

The next step in the study of the evolution of our dynamical system is to find its fixed (or critical) points $(x^{1\star},x^{2\star},\dots, x^{5\star})$, which are given by the conditions
\begin{equation} f^i(x^{1\star},x^{2\star},\dots, x^{5\star})=0.
\end{equation} The stability of the fixed points $(x^{1\star},x^{2\star},\dots, x^{5\star})$ is then analyzed by studying the linearized dynamical system obtained by expanding the evolution equations about those fixed points (as explained in many seminal references, e.g \cite{hirsch}). After that, one tries solutions in the form $(x^1,x^2,\dots,x^5)=(c_1,c_2,\dots,c_5)\,e^{\lambda t}$ in the linear approximation, and finds that their characteristic exponent $\lambda$ and the constant vector $(c_1,c_2,\dots,c_5)$ must be respectively an eigenvalue and an eigenvector of the matrix

\begin{equation}\\
\mathbf{A}=\left( \begin{array}{cccc} \displaystyle\frac{\partial
{x^1}\hspace{1pt}'}{\partial x^1}& \displaystyle\frac{\partial
{x^1}\hspace{1pt}'}{\partial x^2}& \dots&
\displaystyle\frac{\partial {x^1}\hspace{1pt}'}{\partial x^5}\\
\vspace{-8pt}\\
   \displaystyle\frac{\partial {x^2}\hspace{1pt}'}{\partial x^1}&
\displaystyle\frac{\partial {x^2}\hspace{1pt}'}{\partial x^2}&
\dots&
\displaystyle\frac{\partial {x^2}\hspace{1pt}'}{\partial x^5}\\
\vspace{-8pt}\\
\vdots&\vdots&\dots&\vdots\\
\displaystyle\frac{\partial {x^5}\hspace{1pt}'}{\partial
x^1}&\displaystyle\frac{\partial {x^5}\hspace{1pt}'}{\partial
x^2}&\dots&\displaystyle\frac{\partial
{x^5}\hspace{1pt}'}{\partial x^5}
\end{array} \right)_{(x^{1},x^{2},\dots, x^{5})=(x^{1\star},x^{2\star},\dots, x^{5\star})}\\
.\vspace{5pt}\\
\end{equation}

The character of the fixed points depends on the values of the characteristic exponents:
if the real part of all characteristic exponents is negative, the fixed point is asymptotically stable, i.e., an attractor. On the other hand, it is enough to have (at least) one characteristic exponent with positive real part to make the fixed point asymptotically unstable (commonly called as source or repeller). The repeller is a saddle point if at least one of the other characteristic exponents has a negative real part, in which case, apart from the unstable manifold, there is a stable manifold containing the exceptional orbits that converge to the fixed point. 

In addition, when one of the exponents is null the point is not hyperbolic and therefore structural stability cannot be guaranteed (the geometric form of the trajectories may change under small perturbations). Hence, the case in which the largest real part is precisely zero must be analyzed  using other methods. In this case the linear analysis is inconclusive (the Hartman-Grobman theorem fails).

\end{document}